\documentclass[letterpaper, 10 pt, conference]{ieeeconf}

\usepackage{amsmath,amssymb,bm,graphicx,multirow,todonotes}
\usepackage[caption=false,font=footnotesize]{subfig}

\IEEEoverridecommandlockouts
\title{\LARGE \bf
Fast and Robust 3-D Sound Source Localization with DSVD-PHAT
}

\author{Fran\c{c}ois Grondin and James Glass
\thanks{This work was supported in part by the Toyota Research Institute and by the Fonds de recherche du Qu\'{e}bec -- Nature et technologies.}
\thanks{F. Grondin and J. Glass are with the Computer Science \& Artificial Intelligence Laboratory (CSAIL), Massachusetts Institute of Technology, Cambridge, MA 02139, USA \scriptsize\texttt{\{fgrondin,glass\}@mit.edu}}}

\DeclareMathOperator*{\argmax}{arg\,max}
\DeclareMathOperator*{\argmin}{arg\,min}
\DeclareMathOperator{\Tr}{Tr}

\begin{document}

\maketitle
\thispagestyle{empty}
\pagestyle{empty}

\begin{abstract}
This paper introduces a variant of the Singular Value Decomposition with Phase Transform (SVD-PHAT), named Difference SVD-PHAT (DSVD-PHAT), to achieve robust Sound Source Localization (SSL) in noisy conditions.
Experiments are performed on a Baxter robot with a four-microphone planar array mounted on its head.
Results show that this method offers similar robustness to noise as the state-of-the-art Multiple Signal Classification based on Generalized Singular Value Decomposition (GSVD-MUSIC) method, and considerably reduces the computational load by a factor of 250.
This performance gain thus makes DSVD-PHAT appealing for real-time application on robots with limited on-board computing power.
\end{abstract}

\section{INTRODUCTION}

Robot audition aims to provide robots with hearing capabilities to interact efficiently with people in everyday environments \cite{okuno2004computational}.
Sound source localization (SSL) is a typical task that consists of localizing the direction of arrival (DOA) of a target source using a microphone array.
This task is challenging as the robot usually generates a significant amount of noise (fans, actuators, etc.) \cite{ince2010robust} and the target sound source is corrupted by reverberation.
SSL often relies on Multiple Signal Classification (MUSIC) and Steered-Response Power Phase Transform (SRP-PHAT) methods.

MUSIC is a localization method based on Standard Eigenvalue Decomposition (SEVD-MUSIC) that was initially used for narrowband signals \cite{schmidt1986multiple}, and then adapted to broadband signals like speech \cite{ishi2009evaluation}.
However, SEVD-MUSIC assumes the speech signal is more powerful than noise at each frequency bin in the spectrogram, which is usually not the case.
To cope with this limitation, Nakamura et al. introduced the MUSIC based on Generalized Eigenvalue Decomposition (GEVD-MUSIC) method \cite{nakamura2009intelligent, nakamura2011intelligent, nakadai2012robot}.
This method solves the limitation of SEVD-MUSIC, but also introduces some localization errors because the transform provides a noise subspace with correlated bases.
To deal with this issue, a variant of GEVD-MUSIC, named MUSIC based on Generalized Singular Value Decomposition (GSVD-MUSIC), enforces orthogonality between the noise subspace bases and thus improves the DOA estimation accuracy \cite{nakamura2012real}.
However, all MUSIC-based methods rely on online eigenvalue or singular value decompositions that are computationally expensive, and make on-board real-time processing challenging \cite{ohata2014improvement}.

SRP-PHAT is built on the Generalized Cross-Correlation with Phase Transform (GCC-PHAT) between each pair of microphones \cite{brandstein1997robust}.
GCC-PHAT is often computed with the Inverse Fast Fourier Transform (IFFT) to speed up computation, at the cost of discretizing Time Difference of Arrival (TDOA) values, which reduces localization accuracy.
SRP-PHAT usually scans a discretized 3-D space and returns the most likely DOA \cite{valin2003robust,valin2004localization,valin2006robust,valin2007robust,badali2009evaluating,grondin2013manyears}.
This scanning process often involves a significant amount of lookups in memory, which creates a bottleneck and increases execution time.
To reduce the number of lookups, a hierarchical search is proposed to speed up the space scan, but this method still relies on discrete TDOA \cite{grondin2019lightweight}.
We therefore recently proposed the Singular Value Decomposition with Phase Transform (SVD-PHAT) method, which avoids TDOA discretization, and significantly reduces computing time \cite{grondin2019svdphat}.
However, as for SRP-PHAT, SVD-PHAT remains sensitive to additive noise.
To cope with this limitation, time-frequency (TF) masks can be generated to improve robustness to stationary noise \cite{grondin2015time,grondin2016noise}.
Stationary noise is often estimated with techniques like Minima Controlled Recursive Averaging (MCRA) \cite{cohen2002noise} and Histogram-based Recursive Level Estimation (HRLE) \cite{nakajima2010easily}, or recorded offline prior to test if the robot's environment is static.
Pertil\"a et al. also propose a method that generates TF masks using convolutional neural networks for non-stationary noise sources \cite{pertila2017robust}.
However, these TF masks ignore noise spatial coherence, which carries useful insights for robust localization, and is in fact exploited by GSVD-MUSIC.

In this paper, we propose a variant of the SVD-PHAT method, called Difference SVD-PHAT (DSVD-PHAT), that performs correlation matrix subtraction, which considers noise spatial coherence, while preserving the low complexity of the original SVD-PHAT.
Section \ref{sec:gsvdmusic} reviews the state of the art GSVD-MUSIC method, and section \ref{sec:dsvdphat} introduces the proposed DSVD-PHAT method.
Section \ref{sec:setup} describes the experimental setup on a Baxter robot, and then section \ref{sec:results} compares results from GSVD-MUSIC and the proposed DSVD-PHAT approach.

\section{GSVD-MUSIC}
\label{sec:gsvdmusic}

GSVD-MUSIC relies on the Time Difference of Arrival (TDOA) between each microphone and a reference in space.
The TDOA (in sec) stands for the propagation delay for the signal emitted by the sound source DOA $\mathbf{s}_q \in \{\mathbb{R}^3: \lVert \mathbf{s}_q \rVert_2 = 1\}$ (where $\lVert\dots\rVert_2$ stands for the $l_2$-norm) to reach microphone $\mathbf{r}_m \in \mathbb{R}^3$ with respect to the origin.
For discrete-time signals, the TDOA is usually expressed in terms of samples, as shown in (\ref{eq:gsvdmusic_tdoa}), where $c \in \mathbb{R}^+$ stands for the speed of sound in air (in m/sec), and $f_S \in \mathbb{R}^+$ is the sample rate (in samples/sec).
The operator $\cdot$ stands for the dot product.
\begin{equation}
    \tau_{q,m} = \left(\frac{f_S}{c}\right)\mathbf{r}_m \cdot \mathbf{s}_q
    \label{eq:gsvdmusic_tdoa}
\end{equation}

The expression $X^{l}_{m}[k] \in \mathbb{C}$ stands for the Short Time Fourier Transform coefficient of microphone $m \in \{1,\dots,M\}$, at frequency bin $k \in \{0,\dots,N/2\}$ and frame $l \in \mathbb{N}$, where $N \in \mathbb{N}$ and $\Delta N \in \mathbb{N}$ stand for the frame and hop sizes in samples, respectively.
The STFT values are concatenated in the vector $\mathbf{x}^{l}[k] \in \mathbb{C}^{M \times 1}$, as shown in (\ref{eq:gsvdmusic_xv}).
\begin{equation}
    \mathbf{x}^l[k] = \left[
        \begin{array}{cccc}
            X^l_{1}[k] & X^l_{2}[k] & \cdots & X^l_{M}[k] \\
        \end{array}
    \right]^T
    \label{eq:gsvdmusic_xv}
\end{equation}

GSVD-MUSIC uses a steering vector $\mathbf{A}_{q}[k] \in \mathbb{C}^{M\times 1}$ for each potential DOA $\mathbf{s}_q$:
\begin{equation}
    \mathbf{A}_{q}[k] = \left[
        \begin{array}{ccc}
            A_{q,1}[k] & \cdots & A_{q,M}[k] \\
        \end{array}
    \right]^T
    \label{eq:gsvdmusic_Av}
\end{equation}
where $A_{q,m}[k] = \exp{(-2\pi\sqrt{-1}k\tau_{q,m}/N)}$.

The $\mathbb{C}^{M\times M}$ correlation matrix of the vector $\mathbf{x}^{l}[k]$ at each frequency bin $k$ can be estimated at each frame $l$ using the following recursive approximation, where the parameter $\alpha \in\ (0,1)$ is the adaptive rate:
\begin{equation}
    \mathbf{R}_{xx}^l[k] = (1-\alpha)\mathbf{R}_{xx}^{l-1}[k] + \alpha \mathbf{x}^l[k](\mathbf{x}^l[k])^H
    \label{eq:gsvdmusic_Rxx}
\end{equation}
where $\{\dots\}^H$ stands for the Hermitian operator.

The GSVD-MUSIC method performs a generalized singular value decomposition with respect to the noise correlation matrix $\mathbf{R}_{nn}^l[k]$ (which can be estimated as in (\ref{eq:gsvdmusic_Rxx}) during silence periods or precomputed offline if the test environment is known):
\begin{equation}
    (\mathbf{R}_{nn}^l[k])^{-1}\mathbf{R}_{xx}^l[k] = \mathbf{E}^{l}[k]\bm{\Lambda}^{l}[k](\mathbf{F}^{l}[k])^H
    \label{eq:gsvdmusic_gsvd}
\end{equation}
where the diagonal matrix $\bm{\Lambda}^{l}[k] \in (\mathbb{R}^+)^{M\times M}$ holds the singular values in descending order ($\lambda_{1}^{l}[k] > \lambda_{2}^{l}[k] > \dots > \lambda_{M}^{l}[k]$), and $\mathbf{E}^{l}[k] \in \mathbb{C}^{M\times M}$ and $\mathbf{F}^{l}[k] \in \mathbb{C}^{M\times M}$ are the left and right singular vectors $\mathbf{e}_{1}^{l}[k], \dots, \mathbf{e}_{M}^{l}[k]  \in \mathbb{C}^{M\times 1}$ and $\mathbf{f}_{1}^{l}[k], \dots, \mathbf{f}_{M}^{l}[k]  \in \mathbb{C}^{M\times 1}$, respectively:
\begin{equation}
    \bm{\Lambda}^{l}[k] = \left[
    \begin{array}{ccc}
    \lambda_{1}^{l}[k] & \dots & 0 \\
    \vdots & \ddots & \vdots \\
    0 & \dots & \lambda_{M}^{l}[k] \\
    \end{array}
    \right]
    \label{eq:gsvdmusic_Lambda}
\end{equation}
\begin{equation}
    \mathbf{E}^{l}[k] = \left[\mathbf{e}_{1}^{l}[k],\ \dots,\  \mathbf{e}_{M}^{l}[k]\right]
    \label{eq:gsvdmusic_E}
\end{equation}
\begin{equation}
    \mathbf{F}^{l}[k] = \left[\mathbf{f}_{1}^{l}[k],\ \dots,\  \mathbf{f}_{M}^{l}[k]\right]
    \label{eq:gsvdmusic_F}
\end{equation}

This method projects the steering vector $\mathbf{A}_{q}[k]$ in the noise subspace, spanned by the singular vectors $\mathbf{e}^{l}_{m}[k]\ \forall\ m \in \{2, 3, \dots, M\}$ (when there is only one target source).
The inverse of the projections for each frequency bin $k$ is summed over the full spectrum (which may also be restricted to a more specific range of frequency bins \cite{nakamura2012real}):
\begin{equation}
    P_{q}^{l} = \sum_{k=0}^{N/2}{\left(\sum_{m=2}^{M}{\lVert(\mathbf{A}_{q}[k])^H\mathbf{e}_{m}^{l}[k]\rVert_2}\right)^{-1}}
    \label{eq:gsvdmusic_P}
\end{equation}

The sound source DOA then corresponds to $\mathbf{s}_{\bar{q}_l}$, where:
\begin{equation}
    \bar{q}_l = \argmax_{q}{\{P_{q}^{l}\}}
    \label{eq:gsvdmusic_ql}
\end{equation}

GSVD-MUSIC involves $(N/2+1)$ singular value decompositions of $M \times M$ matrices per frame, as shown in (\ref{eq:gsvdmusic_gsvd}), which is challenging from a computing point of view for real-time applications.
Moreover, it also involves computing (\ref{eq:gsvdmusic_P}) for $Q$ potential sources, which also implies a significant amount of computations.
The proposed DSVD-PHAT aims to reduce the amount of computations, while preserving a similar robustness to noise.

\section{DSVD-PHAT}
\label{sec:dsvdphat}

DSVD-PHAT relies on the TDOA between each pair of microphones $i$ and $j$ (as opposed to (\ref{eq:gsvdmusic_tdoa}), where the TDOA is between a microphone and to the origin), which leads to the following expression, for a total of $P=M(M-1)/2$ pairs:
\begin{equation}
    \tau_{q,i,j} = \frac{f_S}{c}(\mathbf{r}_j - \mathbf{r}_i) \cdot \mathbf{s}_q
    \label{eq:dsvdphat_tdoa}
\end{equation}

Since noise and speech sources are independent, it is reasonable to assume that the clean speech correlation matrix $\mathbf{R}_{ss}^{l}$ can be estimated from the difference between the noisy speech and the noise correlation matrices at each frame $l$, as proposed in \cite{higuchi2016robust}:
\begin{equation}
    \mathbf{R}_{ss}^{l}[k] \approx \mathbf{R}_{xx}^{l}[k] - \mathbf{R}_{nn}^{l}[k]
    \label{eq:dsvdphat_Rss}
\end{equation}

The normalized cross-spectra in DSVD-PHAT at each frequency bin $k$ are thus obtained as follows, where $(\dots)_{i,j}$ refers to the element in the $i$th row and $j$th column:
\begin{equation}
    X^{l}_{i,j}[k] = \frac{(\mathbf{R}_{ss}^{l}[k])_{i,j}}{\lVert(\mathbf{R}_{ss}^{l}[k])_{i,j}\rVert_2}
    \label{eq:dsvdphat_X}
\end{equation}

Note how DSVD-PHAT differs from the original SVD-PHAT, as the latter uses directly the noisy correlation matrix (e.g. $\mathbf{R}_{xx}^{l}[k]$ replaces $\mathbf{R}_{ss}^{l}[k]$ in (\ref{eq:dsvdphat_Rss})).

We then define the vector $\mathbf{X} \in \mathbb{C}^{P(N/2+1)\times 1}$ to concatenate all normalized cross-spectra introduced in (\ref{eq:dsvdphat_X}):
\begin{equation}
    \mathbf{X}^l = \left[
        \begin{array}{cccc}
            X^l_{1,2}[0] & X^l_{1,2}[1] & \cdots & X^l_{M-1,M}[N/2] \\
        \end{array}
    \right]^T
    \label{eq:dsvdphat_Xv}
\end{equation}

The matrix $\mathbf{W} \in \mathbb{C}^{Q\times P(N/2+1)}$ holds all the SRP-PHAT coefficients $W_{q,i,j}[k] = \exp{(2\pi\sqrt{-1} k\tau_{q,i,j}/N)}$ :
\begin{equation}
\mathbf{W} = \left[
\begin{array}{cccc}
    W_{1,1,2}[0] & W_{1,1,2}[1] & \cdots & W_{1,M-1,M}[N/2] \\
    \vdots & \vdots & \ddots & \vdots \\
    W_{Q,1,2}[0] & W_{Q,1,2}[1] & \cdots & W_{Q,M-1,M}[N/2] \\
\end{array}
\right]
\label{eq:dsvdphat_W}
\end{equation}

The vector $\mathbf{Y}^l \in \mathbb{R}^{Q\times 1}$ stores the SRP-PHAT energy for all $Q$ potential DOAs, where $\Re\{\dots\}$ extracts the real part of the expression:
\begin{equation}
    \mathbf{Y}^l = \left[
        \begin{array}{ccc}
            Y_1^l & \dots & Y_Q^l
        \end{array}
    \right]^T = \Re\{\mathbf{W}\mathbf{X}^l\}
    \label{eq:dsvdphat_Yv}
\end{equation}

The sound source DOA corresponds to $\mathbf{s}_{\bar{q}_l}$, where:
\begin{equation}
    \bar{q}_l = \argmax_{q}{\{Y_{q}^{l}\}}
    \label{eq:dsvdphat_ql_max}
\end{equation}

Computing $Y^l_q$ for all values of $q$ is expensive, and therefore SVD-PHAT provides a more efficient way of finding $\bar{q}_l$.
The Singular Value Decomposition is first performed on the $\mathbf{W}$ matrix, where $\mathbf{U} \in \mathbb{C}^{Q \times K}$, $\mathbf{S} \in \mathbb{C}^{K \times K}$ and $\mathbf{V} \in \mathbb{C}^{P(N/2+1) \times K}$:
\begin{equation}
    \mathbf{W} \approx  \mathbf{U}\mathbf{S}\mathbf{V}^H
    \label{eq:dsvdphat_West}
\end{equation}

The parameter $K \in \{1, 2, \dots, K_{max}\}$ (where $K_{max} = \max\{Q,P(N/2+1)\}$) satisfies the condition in (\ref{eq:dsvdphat_tr}), which ensures accurate reconstruction of $\mathbf{W}$, where $\delta \in (0,1)$ is a user-defined small value that stands for the tolerable reconstruction error.
The operator $\Tr\{\dots\}$ represents the trace of the matrix.
\begin{equation}
    \Tr{\{\mathbf{S}\mathbf{S}^T\}} \geq (1 - \delta)\Tr{\{\mathbf{W}\mathbf{W}^H\}}
    \label{eq:dsvdphat_tr}
\end{equation}

The vector $\mathbf{Z}^{l} \in \mathbb{C}^{K \times 1}$ results from the projection of the observations $\mathbf{X}^{l}$ in the K-dimensions subspace:
\begin{equation}
    \mathbf{Z}^l = \mathbf{V}^H\mathbf{X}^l
    \label{eq:dsvdphat_Z}
\end{equation}

Similarly, the matrix $\mathbf{D} \in \mathbb{C}^{Q \times K}$ holds a set of $Q$ vectors $\mathbf{D}_q \in \mathbb{C}^{1\times K}$:
\begin{equation}
    \mathbf{D} = \mathbf{U}\mathbf{S} = \left[
    \begin{array}{cccc}
        \mathbf{D}_1^T & \mathbf{D}_2^T & \dots & \mathbf{D}_Q^T \\
    \end{array}
    \right]^T
    \label{eq:dsvdphat_D}
\end{equation}

The optimization in (\ref{eq:dsvdphat_ql_max}) can then be converted to a nearest neighbor problem:
\begin{equation}
    \bar{q}_l = \argmin_{q}{\{\lVert \hat{\mathbf{D}}_q - (\hat{\mathbf{Z}}^l)^H \rVert^2_2\}}
    \label{eq:dsvdphat_ql_min}
\end{equation}
where $\hat{\mathbf{D}}_q = \mathbf{D}_{q}/\lVert \mathbf{D} \rVert_2$ and $\hat{\mathbf{Z}}_q = \mathbf{Z}_{q}/\lVert \mathbf{Z} \rVert_2$.
A k-d tree then solves efficiently this nearest neighbor search problem.
The corresponding amplitude for the optimal DOA at index $\bar{q}_l$ corresponds to:
\begin{equation}
    Y^l_{\bar{q}_l} = \mathbf{W}_{\bar{q}_l}\mathbf{X}^l
    \label{eq:dsvdphat_Yq}
\end{equation}
where $\mathbf{W}_{\bar{q}_l}$ stands for the $\bar{q}_l$-th row of $\mathbf{W}$.

Both GSVD-MUSIC and DSVD-PHAT rely on SVD decompositions, but DSVD-PHAT computes them offline.
The online processing only involves the projection in (\ref{eq:dsvdphat_Z}) and the k-d tree search, which is appealing for real-time processing.

\section{EXPERIMENTAL SETUP}
\label{sec:setup}

\begin{figure}[!ht]
    \centering
    \includegraphics[width=0.8\columnwidth]{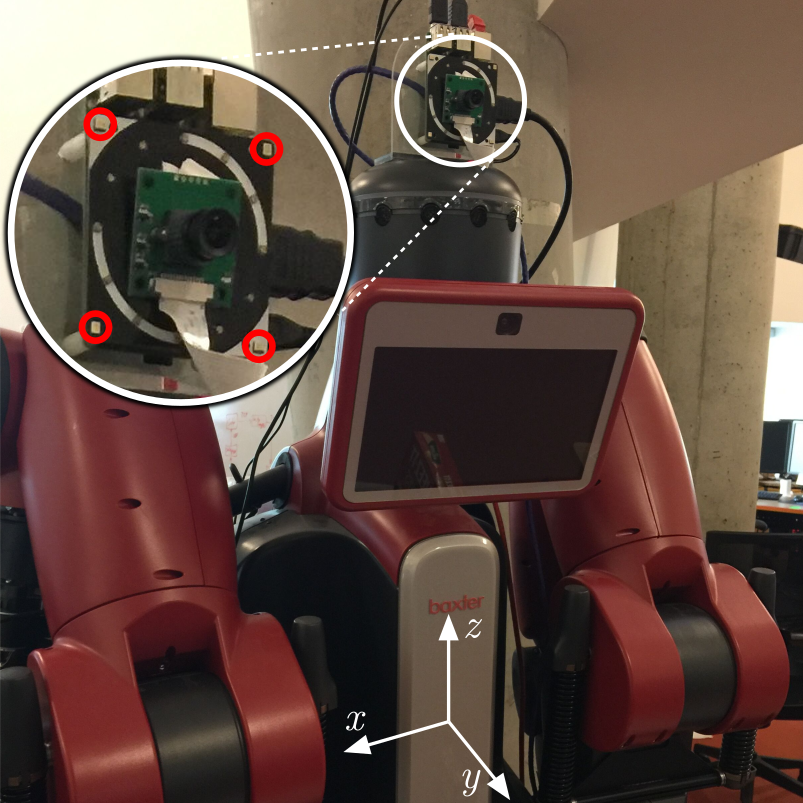}
    \caption{Baxter robot equipped with a 4-microphone ReSpeaker array mounted on its head (microphones are circled in red)}
    \label{fig:setup_baxter}
\end{figure}

The GSVD-MUSIC and DSVD-PHAT methods are evaluated for a Baxter robot setup, equipped with a 4-microphone ReSpeaker\footnote{\texttt{http://seeedstudio.io}} array mounted on its head, as shown in Fig. \ref{fig:setup_baxter}.

To compare both methods with a wide range of conditions, we perform simulations to evaluate numerous room configurations and signal-to-noise ratios (SNRs).
Noise from Baxter's fans is therefore recorded and then mixed with male and female speech utterances from the TIMIT dataset \cite{zue1990speech}, convolved with simulated Room Impulse Responses (RIRs) and amplified with various gains.
The room impulse response (RIR) corresponds to the impulse response obtained with the image method \cite{allen1979image} between the microphone array and the target sound sources, both positioned randomly in a $10$m x $10$m x $3$m room.
For each pair of SNR and room reverberation time RT60, we generate $100$ RIRs and use the same number of speech sources picked randomly from the TIMIT dataset.

The parameters for the experiments are summarized in Table \ref{tab:setup_parameters}.
The sample rate $f_S$ captures all the frequency content of speech, and the speed of sound $c$ corresponds to typical indoor conditions.
The frame size $N$ analyzes segments of 16 msecs, and the hop size $\Delta N$ provides a 50\% overlap.
The potential DOAs are represented by equidistant points on a unit halfsphere generated recursively from a tetrahedron, for a total of $1282$ points, as in \cite{grondin2019lightweight}.
The smoothing parameter $\alpha$ provides a context of roughly $800$ msecs to estimate the correlation matrices, which captures multiple phonemes.
The parameter $\delta$ is set to the value found in \cite{grondin2019svdphat}, which ensures a good accuracy.
For this array configuration, the dimensionality of the subspace corresponds to $K = 23$ with $\delta = 10^{-5}$.

\begin{table}[]
    \centering
    \caption{GSVD-MUSIC and GSVD-PHAT Parameters}
    \renewcommand{\arraystretch}{1.3}    
    \begin{tabular}{|cccccccc|}
        \hline
        $f_S$ & $c$ & $M$ & $N$ & $\Delta N$ & $Q$ & $\alpha$ & $\delta$ \\
        \hline
        $16000$ & $343.0$ & $4$ & $256$ & $128$ & $1282$ & $0.05$ & $10^{-5}$ \\
        \hline
    \end{tabular}
    \label{tab:setup_parameters}
\end{table}

Table \ref{tab:setup_mics} lists the positions of the ReSpeaker array microphones (in cm) w.r.t. to the center of the array.

\begin{table}[!ht]
    \centering
    \caption{Positions (x,y,z) of the microphones in cm}
    \renewcommand{\arraystretch}{1.3}
    \begin{tabular}{|c|ccc|}
        \hline
        $m$ & $x$ & $y$ & $z$ \\
        \hline
        $1$ & $+2.9$ & $0.0$ & $+2.9$ \\ 
        $2$ & $+2.9$ & $0.0$ & $-2.9$ \\
        $3$ & $-2.9$ & $0.0$ & $+2.9$ \\
        $4$ & $-2.9$ & $0.0$ & $-2.9$ \\
        \hline
    \end{tabular}
    \label{tab:setup_mics}
\end{table}

In all experiments, the noise correlation matrix comes from the offline recording of the robot's fans.
This ensures we compare both methods independently of the performance of the online background noise estimation method.

\section{RESULTS}
\label{sec:results}

\begin{figure}[!ht]
\centering
  \subfloat[Spectrogram of the signal captured at microphone 1.]{%
  \includegraphics[width=0.99\columnwidth]{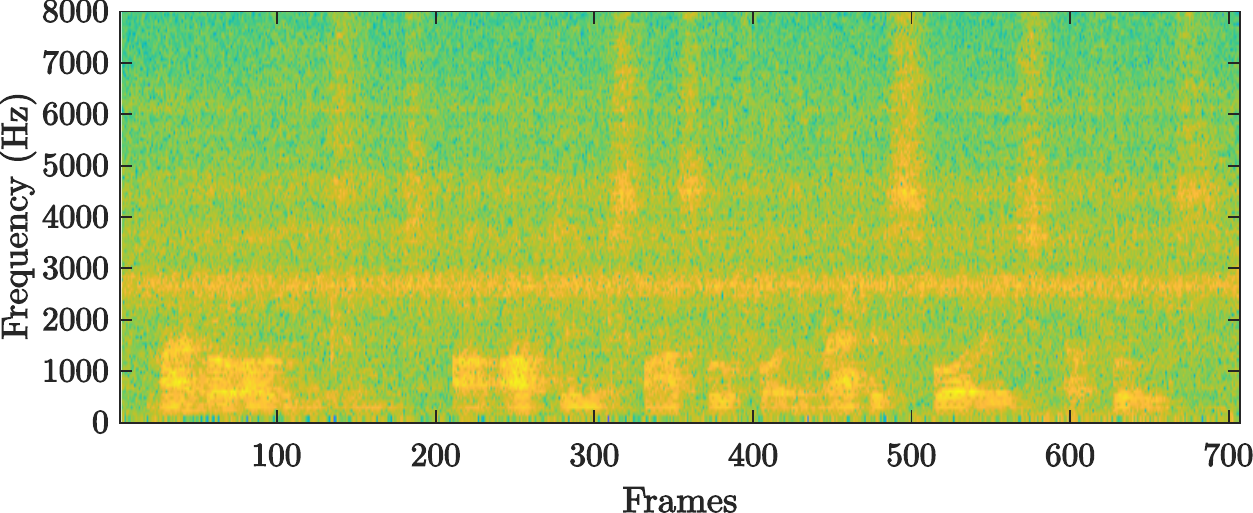} \label{fig:results_ex_01}} \\
  \subfloat[Circles represent the $\mathbf{s}_{\bar{q}_l}$ found with GSVD-MUSIC, and lines stand for the theoretical DOA. The x-, y-, z-coordinates are represented by blue, red and green colors, respectively.]{%
  \includegraphics[width=\columnwidth]{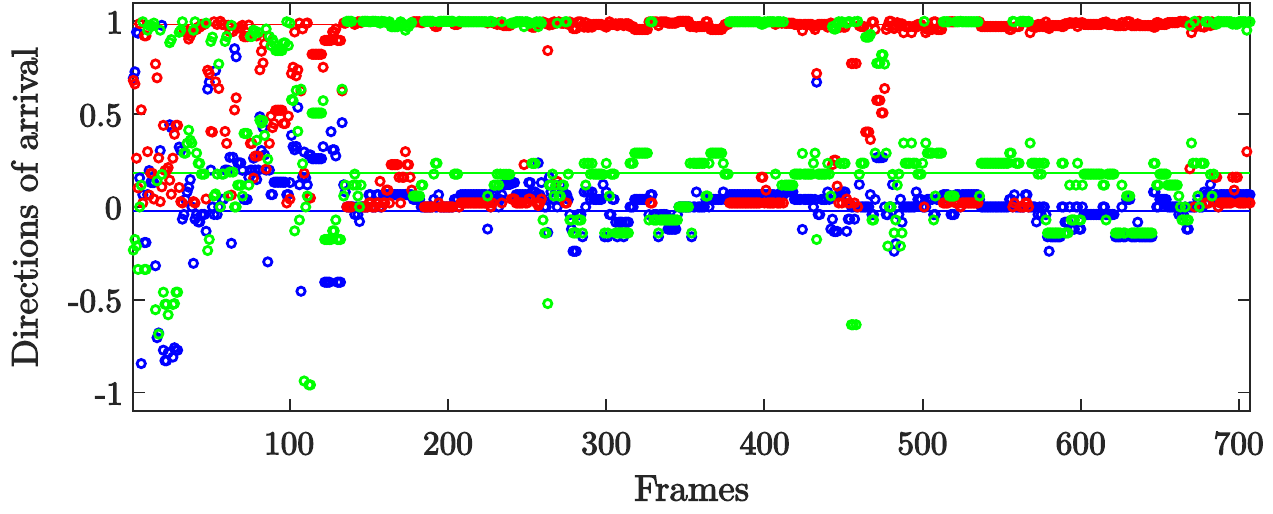}\label{fig:results_ex_02}} \\
  \subfloat[Circles represent the $\mathbf{s}_{\bar{q}_l}$ found with DSVD-PHAT, and lines stand for the theoretical DOA. The x-, y-, z-coordinates are represented by blue, red and green colors, respectively.]{%
  \includegraphics[width=\columnwidth]{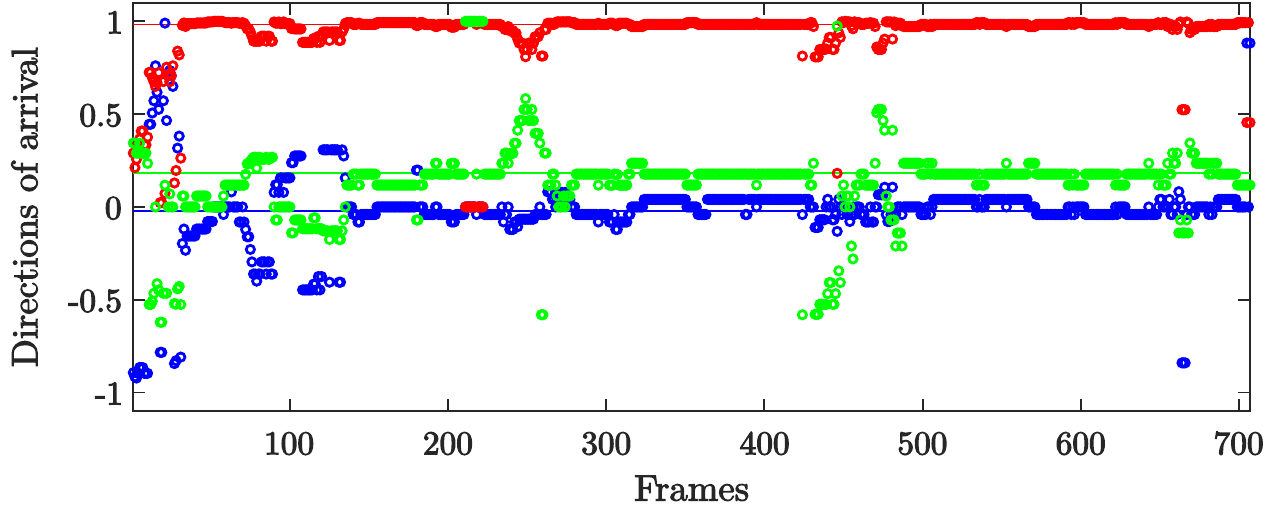}\label{fig:results_ex_03}} \\
\caption{SSL with GSVD-MUSIC and DSVD-PHAT when RT60 = $400$ msecs and SNR = $5$ dB.}
\label{fig:results_ex}
\end{figure}

To get some intuition about the SSL with GSVD-MUSIC and DSVD-PHAT, we first analyze an example of a speech utterance with a SNR of $5$ dB and a reverberation level of RT60 = $400$ msecs, shown in Fig. \ref{fig:results_ex}.
The spectrogram in Fig. \ref{fig:results_ex_01} displays the speech signal, corrupted by some stationary noise between $2500$Hz and $5000$Hz.
Fig. \ref{fig:results_ex_02} shows the DOAs obtained from GSVD-MUSIC, with the true DOA represented by straigh lines.
This example demonstrates that, in this specific case, GSVD-MUSIC estimates many DOAs that differ from the theoretical DOA.
Similarly, Fig. \ref{fig:results_ex_03} displays the DOAs obtained from DSVD-PHAT for the same noisy signal.
Here the estimated DOAs are closer to the theoretical DOA.

It is also convenient to define the expression $\theta_l \in [0,\pi/2]$ to denote the angle difference between the estimated DOA $\mathbf{s}_{q_l}$ at frame $l$ (obtained using GSVD-MUSIC or DSVD-PHAT), and the theoretical DOA $\mathbf{s}_{true}$ extracted from the simulated room parameters:
\begin{equation}
    \theta_l = \arccos{\{\mathbf{s}_{q_l} \cdot \mathbf{s}_{true}\}}
\end{equation}

Let us define the margin $\Delta\theta \in [0,\pi/2]$, that corresponds to the DOA error tolerance for a localized source to be considered as a valid DOA.
In this section, we arbitrary define the tolerance to $\Delta\theta = 0.2 \ \textrm{radians}$, which corresponds to $11.5^{\circ}$.
Expression $\Theta_l$ takes a value of $1$ when the localized sound source is within the range, or $0$ otherwise:
\begin{equation}
    \Theta_l = \begin{cases}
    1 & \theta_l \leq \Delta\theta \\
    0 & \theta_l > \Delta\theta \\
    \end{cases}
\end{equation}

Similarly, the expression $e_l$ corresponds to the observation amplitude ($e_l = P^l_{q_l}$ for GSVD-MUSIC from (\ref{eq:gsvdmusic_P}), and $e_l = Y^l_{q_l}$ from (\ref{eq:dsvdphat_Yq}) for DSVD-PHAT).
This metric is relevant as it is often assumed that the confidence in the DOA $\mathbf{s}_{\bar{q}_l}$ depends on the associated amplitude of $e_l$ \cite{grondin2013manyears, grondin2019lightweight}.
Therefore, a DOA is considered as a positive when the amplitude $e_l$ equals or exceeds the fixed threshold $T_{min}$, and as a negative otherwise:
\begin{equation}
    E_l = \begin{cases}
    1 & e_l \geq T_{min} \\
    0 & e_l < T_{min} \\
    \end{cases}
\end{equation}

\begin{figure}[!ht]
  \centering
  \subfloat[Angle difference $\theta_l$ for GSVD-MUSIC (blue) and $\Delta\theta$ threshold (red).]{%
  \includegraphics[width=\columnwidth]{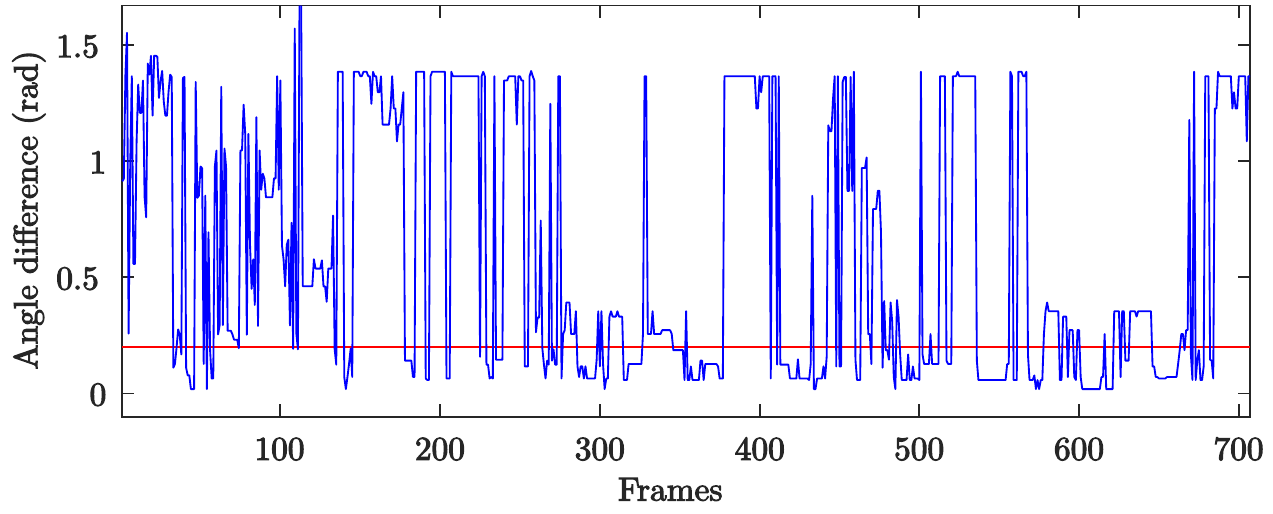}} \\
  \subfloat[Amplitude $e_l = P^l_{\bar{q}_l}$ for GSVD-MUSIC.]{%
  \includegraphics[width=\columnwidth]{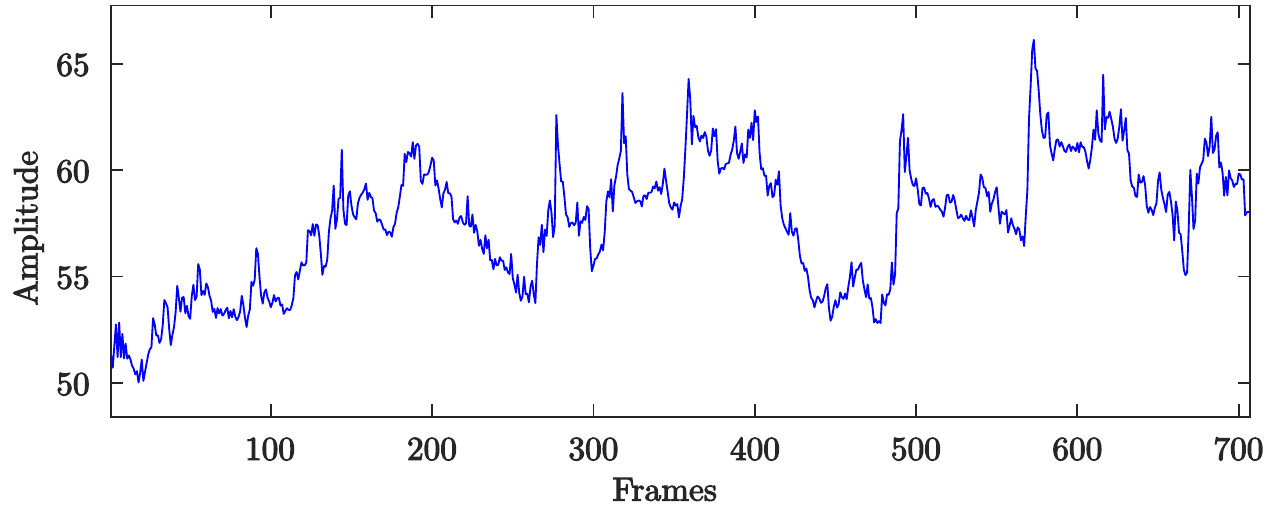}} \\
  \subfloat[Angle difference $\theta_l$ for DSVD-PHAT (blue) and $\Delta\theta$ threshold (red).]{%
  \includegraphics[width=\columnwidth]{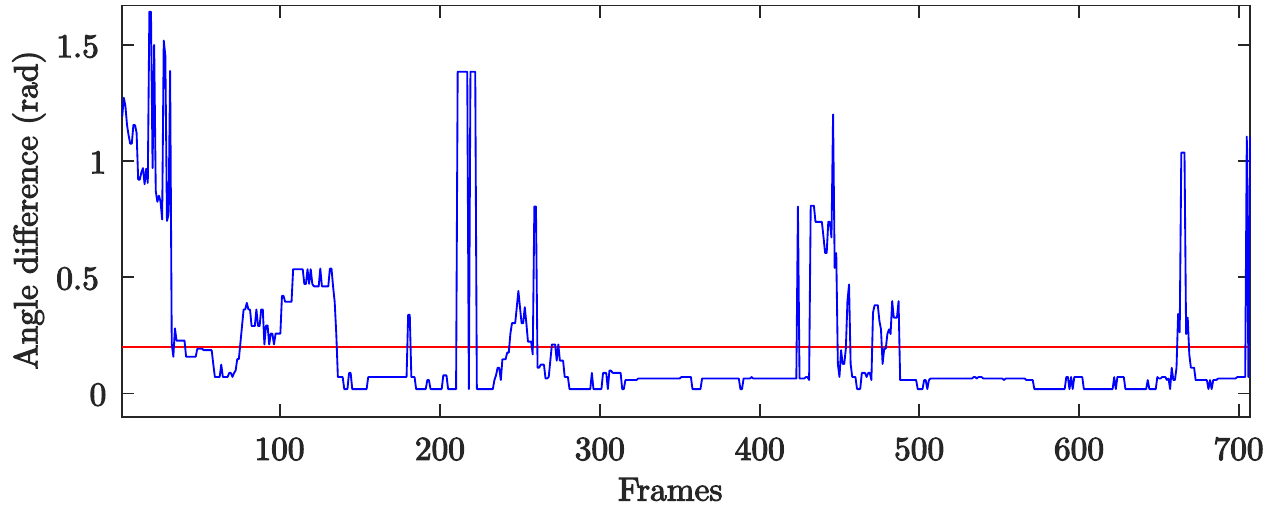}} \\
  \subfloat[Amplitude $e_l = Y^l_{\bar{q}_l}$ for DSVD-PHAT.]{%
  \includegraphics[width=\columnwidth]{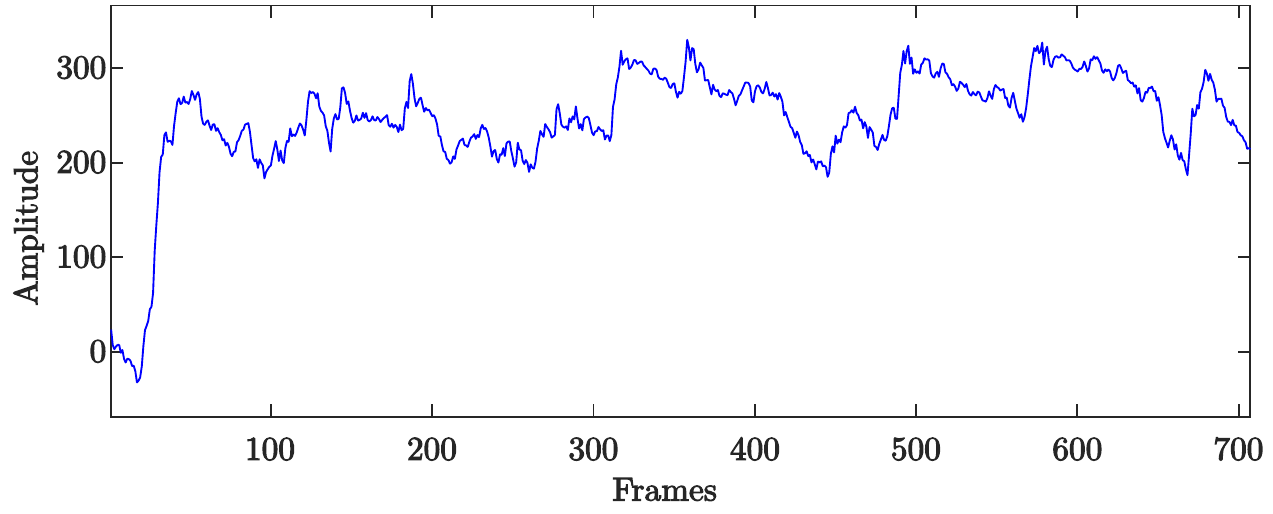}} \\
  \caption{Comparisons between GSVD-MUSIC and DSVD-PHAT methods.} 
  \label{fig:results_thetasamplitudes}
\end{figure}

Fig. \ref{fig:results_thetasamplitudes} illustrates the angle difference of the DOAs estimated previously with both methods, and also displays the associated amplitudes.
Note that for DSVD-PHAT in particular, the amplitude goes down when the value of $\theta$ gets outside the acceptable range, which suggests that a well-tuned $T_{min}$ could discriminate between accurate and inaccurate estimated DOAs.

To measure the performance of both methods, we vary the value of $T_{min}$ and compute the number of true positives (TP), true negatives (TN), false positives (FP) and false negatives (FN).
A TP occurs when the amplitude is greater or equal to the threshold, and the measured DOA falls within the acceptable range of the theoretical DOA:
\begin{equation}
    TP = \sum_{l=0}^{L}{\Theta_l e_l}
\end{equation}

Similarly, a TN happens when a DOA out of the acceptable range is rejected as its associated amplitude is below the fixed threshold:
\begin{equation}
    TN = \sum_{l=0}^{L}{(1-\Theta_l)(1-e_l)}
\end{equation}

Finally, FP and FN occur when an erroneous DOA is picked and when a valid DOA is rejected, respectively:
\begin{equation}
    FP = \sum_{l=0}^{L}{(1-\Theta_l) e_l}
\end{equation}
\begin{equation}
    FN = \sum_{l=0}^{L}{\Theta_l (1-e_l)}
\end{equation}

The True Positive Rate (TPR) and False Positive Rate (FPR) then correspond to (\ref{eq:results_tpr}) and (\ref{eq:results_fpr}), respectively, and are used to build the ROC curve.
\begin{equation}
    TPR = \frac{TP}{TP+FN}
    \label{eq:results_tpr}
\end{equation}
\begin{equation}
    FPR = \frac{FP}{FP+TN}
    \label{eq:results_fpr}
\end{equation}

\begin{figure}[!ht]
    \centering
    \includegraphics[width=0.9\columnwidth]{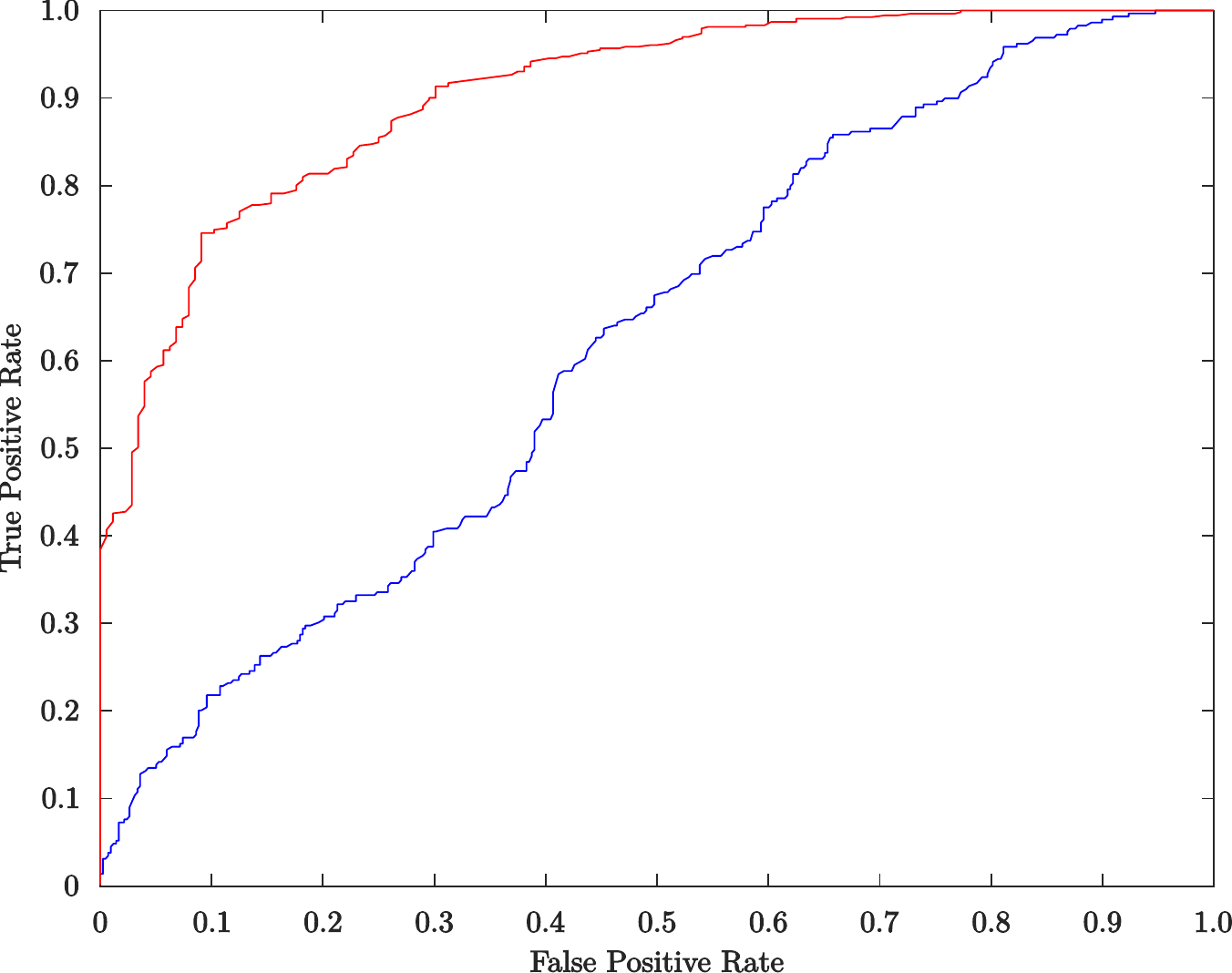}
    \caption{ROC curves for GSVD-MUSIC (blue) and DSVD-PHAT (red).}
    \label{fig:results_auc}
\end{figure}

Fig. \ref{fig:results_auc} shows both ROC curves with GSVD-MUSIC and DSVD-PHAT for the previous example.
In this case, the DSVD-PHAT surpasses the GSVD-MUSIC results as the Area Under the Curve (AUC) is clearly closer to $1$.

Table \ref{tab:my_label} shows the AUC results for SNRs $\in \{-10,-5,\dots,20\}$ dB and RT60 $\in \{200, 400, 600, 800\}$ msecs.
In general, GSVD-MUSIC generates higher AUC values for cases when the SNR is below $0$dB.
However, the DSVD-PHAT still provides AUC values close to GSVD-MUSIC, which demonstrates that the proposed method also allows accurate DOA estimation under reverberant and noisy conditions.
Moreover, the proposed DSVD-PHAT approach provides better results for all scenarios where the SNR is greater or equal to $5$ dB, at all reverberation levels.

\begin{table}[!ht]
    \centering
    \caption{AUC of the ROC Curves}
    \renewcommand{\arraystretch}{1.3}
    \begin{tabular}{|c|c|cc|}
        \hline
        SNR (dB) & RT60 (msec) & GSVD-MUSIC & DSVD-PHAT \\
        \hline
        \multirow{4}{*}{$-10$} & $200$ & $\mathbf{0.68}$ & $0.64$ \\
         & $400$ & $\mathbf{0.55}$ & $0.49$ \\
         & $600$ & $\mathbf{0.52}$ & $0.47$ \\
         & $800$ & $\mathbf{0.51}$ & $0.45$ \\
        \hline
        \multirow{4}{*}{$-5$} & $200$ & $\mathbf{0.77}$ & $0.75$ \\
         & $400$ & $\mathbf{0.66}$ & $0.62$ \\
         & $600$ & $\mathbf{0.59}$ & $0.55$ \\
         & $800$ & $0.52$ & $\mathbf{0.53}$ \\
        \hline
        \multirow{4}{*}{$0$} & $200$ & $\mathbf{0.84}$ & $\mathbf{0.84}$ \\
         & $400$ & $0.70$ & $\mathbf{0.71}$ \\
         & $600$ & $\mathbf{0.66}$ & $0.65$ \\
         & $800$ & $0.63$ & $\mathbf{0.64}$ \\
        \hline
        \multirow{4}{*}{$5$} & $200$ & $0.87$ & $\mathbf{0.91}$ \\
         & $400$ & $0.73$ & $\mathbf{0.76}$ \\
         & $600$ & $0.71$ & $\mathbf{0.74}$ \\
         & $800$ & $\mathbf{0.64}$ & $\mathbf{0.64}$ \\
        \hline        
        \multirow{4}{*}{$10$} & $200$ & $0.93$ & $\mathbf{0.95}$ \\
         & $400$ & $0.76$ & $\mathbf{0.84}$ \\
         & $600$ & $0.71$ & $\mathbf{0.77}$ \\
         & $800$ & $0.64$ & $\mathbf{0.69}$ \\
        \hline
        \multirow{4}{*}{$15$} & $200$ & $0.93$ & $\mathbf{0.98}$ \\
         & $400$ & $0.80$ & $\mathbf{0.86}$ \\
         & $600$ & $0.73$ & $\mathbf{0.80}$ \\
         & $800$ & $0.66$ & $\mathbf{0.69}$ \\
        \hline
        \multirow{4}{*}{$20$} & $200$ & $0.95$ & $\mathbf{0.99}$ \\
         & $400$ & $0.76$ & $\mathbf{0.84}$ \\
         & $600$ & $0.70$ & $\mathbf{0.71}$ \\
         & $800$ & $0.64$ & $\mathbf{0.65}$ \\
        \hline          
    \end{tabular}
    \label{tab:my_label}
\end{table}

Both methods are compared in terms of the execution times per frame.
These methods run in the MATLAB environment, and their implementation relies mostly on vectorization to speed up processing.
The hardware used consists of an Intel Xeon CPU E5-1620 clocked at 3.70GHz.
Table \ref{tab:results_speed} shows the average execution time per frame.
This demonstrates the significant efficiency gain with DSVD-PHAT that avoids the expensive online SVD computations, as it runs approximately $250$ times faster than GSVD-MUSIC.
In this experiment, with $\Delta N/f_S = 8$ msecs between each frame, GSVD-MUSIC requires roughly $300\%$ of the actual computing resources to achieve real-time, whereas DSVD-PHAT easily meets real-time requirements by using only $1\%$ of the computing power.
\begin{table}[!ht]
    \centering
    \caption{EXECUTION TIME PER FRAME}
    \renewcommand{\arraystretch}{1.3}
    \begin{tabular}{|c|cc|}
        \hline
        Method & GSVD-MUSIC & DSVD-PHAT \\
        \hline
        Time (msecs) & 23.3 & 0.093 \\
        \hline
    \end{tabular}
    \label{tab:results_speed}
\end{table}

\section{CONCLUSION}
\label{sec:conclusion}

This paper introduces a variant of the SVD-PHAT method to improve noise robustness.
Results demonstrate that the proposed method performs similarly to the state of the art GSVD-MUSIC technique, but runs approximately $250$ times faster.
This makes DSVD-PHAT appealing for localization on robots with limited on-board computing power.

In future work, we will investigate multiple sound source localization with the proposed DSVD-PHAT method.
Moreover, DSVD-PHAT could be incorporated to existing SSL frameworks such as HARK\footnote{\texttt{http://hark.jp}} \cite{nakadai2010design} and ODAS\footnote{\texttt{http://odas.io}} \cite{grondin2019lightweight}.


\bibliographystyle{IEEEtran}
\bibliography{IEEEabrv,refs}

\end{document}